# Velocity correlation tensor and substructure's statistics around density peaks

**V. Antonuccio-Delogu**[1,3,4] **and S. Colafrancesco**[2,4]

[1] Osservatorio Astrofisico and CNR-GNA, Unità di Ricerca di Catania,Viale A. Doria 6, I-95125 Catania, ITALY
[2] Osservatorio Astronomico di Roma, Via dell'Osservatorio 5, I-00040 Monteporzio (Roma), ITALY
[3] Theoretical Astrophysics Center, Blegdamsvej 17, DK 2100 Copenhagen, DENMARK
[4] Member of the European Cosmology Network ULySSES

**Abstract.** We derive the conditional probability distribution $f_{\rm pk}(v \mid \nu)$ of the peculiar velocity around a peak of given overdensity $\nu$ in a Gaussian density field. This distribution characterizes the shear field around local, isolated density peaks. We use the moments of the velocity distribution to study the spatial distribution of the escape velocity from protostructures forming in a standard, flat, scale–invariant CDM model. In the case of high ($\nu \sim 2$) peaks with filtering scales $R_f \sim 2h^{-1}$ Mpc, we find that at distances between $2 \div 4$ Mpc h$^{-1}$ (or equivalently $\approx 1 \div 2$ $R_f$) a sensitive fraction of the matter enclosed within an extrapolated (up to the present epoch) volume containing a total gravitating mass $M \approx 2.725 \times 10^{15} h^{-1} M_\odot$ (of the same order as that derived from gravitational lensing observations, see Tyson and Fisher 1995) is *gravitationally unbound* to the peak. The fraction of unbound matter varies between approximately between $0.05 - 0.2$ depending on the central height $\nu$ of the density peak, on the filtering radius and on the adopted normalization for the fluctuation spectrum. We show how these results support our previous conclusions regarding the delay of the collapse induced by the presence of small-scale substructure in clusters of galaxies formed in standard CDM cosmological models.

## 1. Introduction

In most of the contemporary, widely studied cosmological scenarios the formation of cosmic structures is traced back to the evolution of primordial density fluctuations. In particular, within the hierarchical clustering scenarios like those envisaged by Cold Dark Matter (hereafter CDM) models, local peaks of the density field are regarded as sites which could have engendered present-day observed structures like galaxies and clusters of galaxies (Kaiser 1986; Bardeen et al. 1986, hereafter BBKS). The structure of the density peaks arising from primordial Gaussian perturbations has been studied by many authors (e.g. Peacock & Heavens 1985; BBKS; González & Martínez 1995), with particular attention to their mass distribution. The structure of the velocity field around a peak, however, has received comparatively lesser attention, although its knowledge can be important for the interpretation of the observed bulk flows (Davis & Peebles 1983; Gorski 1988; Groth et al. 1989) and for a *quantitative* understanding of the collapse and virialization of protostructures (Weygaert & Babul 1994). The first point is particularly important when one compares the predictions of cosmological models with peculiar velocity data. Recently Mo et al. (1993) and Marzke et al. (1995) have shown that the discrepancy beetween the standard CDM predicted galaxy pairwyse velocity dispersion and the observed value is greatly reduced when the contribution from galaxies in rich Abell clusters is subtracted from the samples. It remains to be proved that the *predicted* contribution from these clusters in standard CDM models is of the right order of magnitude to explain the discrepancy. In this paper we will show that this is indeed true, although with a rather large variance.

A knowledge of the velocity distribution function around a density peak can be of importance also to give quantitative information on the amount od substructure around a proto–object. Indeed, in some recent papers (Antonuccio-Delogu, 1992; Antonuccio-Delogu & Colafrancesco 1994; Colafrancesco et al. 1995) we have shown that dynamical friction induced by the presence of *small-scale-substructure* (hereafter *sss*, a natural byproduct of *any* delay of the collapse of the outermost shells of a proto-cluster. However, we also remarked in those papers that only that fraction of the *sss* made of subpeaks *not gravitationally bound* to the perturbation should be considered as part of the "background" and only its contributions should enter into the calculation of the dynamical friction coefficient. The conservative estimate we have adopted for this fraction was $\sim 0.1$. In fact, this fraction can be easily computed once the velocity distribution function is known. As we will see, the calculations we report in this paper will prove that our estimate was indeed a conservative one.

The plan of the paper is the following. In Sec. 2 we will derive the probability distribution of the velocity around peaks of given central overdensity $\nu$, using standard methods from Gaussian random fields' theory (as sketched, e.g. in the Appendix D of BBKS). This distribution is generally anisotropic, and its knowledge fully specifies the velocity field around the peak. In Sec. 3 we will then use this knowledge to determine the *fraction of gravitationally unbound matter*. Finally, in Sec. 4 we will present the conclusions and some hints for future work.

## 2. Velocity probability distribution around a density peak

### 2.1. Conditional probability distributions

Our target is to calculate the *peculiar velocity probability distribution* $f_{\rm pk}(\boldsymbol{v} \mid \nu)$ around a peak of given central overdensity $\nu = \delta_0/\sigma$. The peculiar velocity $\boldsymbol{v}$ is a random variable given (in the linear regime) by:

$$\boldsymbol{v} = -\frac{H_0 \Omega_0^{0.6}}{r^3} \int_0^r dr' r'^2 \bar{\delta}(r') \boldsymbol{r} \tag{1}$$

A standard theorem from the theory of gaussian random fields ensures that, being linearly dependent on a gaussian random variable (the overdensity $\delta(r)$), the peculiar velocity will also be randomly gaussian distributed.

We consider here relatively isolated, high $\nu$ density peaks which are nearly spherically symmetric. The analogous probability distribution around *two* peaks of the density field has been recently considered by Regõs and Szalay (1995).

We will first compute a more general distribution: $f(\boldsymbol{v}, x \mid \nu, \boldsymbol{\eta})$, i.e. the conditional probability for velocity *and* central curvature ($x \equiv -\nabla^2 \delta/\sigma_2 \mid_0$) for peaks of given $\nu$ whose central shape is described by the gradient of the field, $\boldsymbol{\eta} = <\nabla \delta(r)>$, where $\delta(r)$ is the density field found around the peak (see e.g. BBKS).

We will then average this probability distribution over **maxima** ($x > 0$) imposing the condition $\boldsymbol{\eta} \equiv 0$, characterizing points of the random process that are extrema. Note that $f(\boldsymbol{v}, x \mid \nu, \boldsymbol{\eta})$ denotes the probability distribution of the 4-vector $(\boldsymbol{v}, x)$ given that the quantities $(\nu, \boldsymbol{\eta})$

The above probability distribution is a conditional multivariate probability (under the hypothesys that all the above variables are Gaussian distributed), and can be computed using standard theorems. Let us define $\boldsymbol{Y}_{\rm A} \equiv (\nu, \boldsymbol{\eta})$, $\boldsymbol{Y}_{\rm B} \equiv (\boldsymbol{v}, x)$. If $\boldsymbol{Y}_{\rm A}, \boldsymbol{Y}_{\rm B}$ are Gaussian distributed, then the multivariate probability distribution for the vector $\boldsymbol{Y}_{\rm B}$ given that the vector $\boldsymbol{Y}_{\rm A}$ has a given value, is a Gaussian with mean:

$$\langle \boldsymbol{Y}_{\rm B} \mid \boldsymbol{Y}_{\rm A} \rangle = \langle \boldsymbol{Y}_{\rm B} \otimes \boldsymbol{Y}_{\rm A} \rangle \langle \boldsymbol{Y}_{\rm A} \otimes \boldsymbol{Y}_{\rm A} \rangle^{-1} \boldsymbol{Y}_{\rm A}^{\dagger} \tag{2}$$

and covariance matrix:

$$\langle \Delta \boldsymbol{Y}_{\rm B} \otimes \Delta \boldsymbol{Y}_{\rm B} \mid \boldsymbol{Y}_{\rm A} \rangle =$$

$$\langle \boldsymbol{Y}_{\rm B} \otimes \boldsymbol{Y}_{\rm B} \rangle - \langle \boldsymbol{Y}_{\rm B} \otimes \boldsymbol{Y}_{\rm A} \rangle \langle \boldsymbol{Y}_{\rm A} \otimes \boldsymbol{Y}_{\rm A} \rangle^{-1} \langle \boldsymbol{Y}_{\rm A} \otimes \boldsymbol{Y}_{\rm B} \rangle \tag{3}$$

(see BBKS, Appendix D). Here $\Delta \boldsymbol{Y}_{\rm B} = \boldsymbol{Y}_{\rm B} - \langle \boldsymbol{Y}_{\rm B} \mid \boldsymbol{Y}_{\rm A} \rangle$, and the symbol $\otimes$ denotes a matrix product.

The explicit calculation of these matrices is not difficult but a little tedious and can be easily done once some averages like $\langle \boldsymbol{v}\nu \rangle$ are computed (to make easier the job to the interested reader we calculate all the relevant non-zero averages in the Appendix A). It turns out that the vector $\Delta \boldsymbol{Y}_{\rm B}$ is given by:

$$\Delta Y_{\rm B(i)} \equiv \begin{pmatrix} v_1 - \nu \langle v_1 \nu \rangle - \frac{3}{\sigma_1^2} \sum_{i=1}^{3} \eta_i \langle v_1 \eta_i \rangle \\ v_2 - \nu \langle v_2 \nu \rangle - \frac{3}{\sigma_1^2} \sum_{i=1}^{3} \eta_i \langle v_2 \eta_i \rangle \\ v_3 - \nu \langle v_3 \nu \rangle - \frac{3}{\sigma_1^2} \sum_{i=1}^{3} \eta_i \langle v_3 \eta_i \rangle \\ x - \nu \langle \nu x \rangle \end{pmatrix} \tag{4}$$

and the final result for maxima ($\boldsymbol{\eta} = 0$, $x > 0$) reads:

$$f(\boldsymbol{v}, x \mid \nu, \boldsymbol{\eta} = 0) = \frac{1}{(2\pi)^2 (\prod_{i=1}^{4} \lambda_i)^{1/2}} \cdot$$

$$\cdot \exp \left( -\sum_{i=1}^{3} \frac{\Delta Y_{B(i)}^2}{\lambda_i} - \frac{[x - \nu \langle \nu x \rangle]^2}{\lambda_4} \right) \tag{5}$$

In this latter equation we have introduced the eigenvalues $\lambda_i, i = 1, ..4$ of the covariance matrix $\mathsf{M} \equiv \langle \Delta \boldsymbol{Y}_{\rm B} \otimes \Delta \boldsymbol{Y}_{\rm A} \rangle$ whose elements $M_{ij}$ are:

$$M_{ii} = \Pi(r) - \left\{ \langle v_i \nu \rangle^2 + \frac{3}{\sigma_1^2} \sum_{j=1}^{3} \langle v_i \eta_j \rangle \right\}, \; i = 1, ...3$$

$$M_{44} = \langle \nu^2 \rangle - \langle \nu x \rangle^2$$

$$M_{ij} = \Pi(r) - \Sigma(r) - \left\{ \langle v_i \nu \rangle \langle v_j \nu \rangle + \frac{3}{\sigma_1^2} \sum_{l=1}^{3} \langle v_i \eta_l \rangle \langle v_j \eta_l \rangle \right\},$$

$$i, j = 1, ...3, i \neq j$$

$$M_{i4} = \langle v_i x \rangle - \langle v_i \nu \rangle \langle \nu x \rangle, \; i = 1, ...3 \tag{6}$$

functions $\Pi(r), \Sigma(r)$ are the parallel and transverse components of the velocity tensor for the background field:

$$\langle \mathbf{v}(\mathbf{r}')\mathbf{v}(\mathbf{r}'+\mathbf{r})\rangle \equiv [\Pi(r) - \Sigma(r)]\hat{r}_\alpha \hat{r}_\beta + \Sigma(r)\delta_{\alpha\beta} \qquad (7)$$

These functions enter our formula because of the relation $\langle v_i v_j \rangle(\mathbf{r}) = \langle v_i(\mathbf{r}')v_j(\mathbf{r}'+\mathbf{r})\rangle$ holding in an homogeneous background.

Formulas for $\Pi(r)$ and $\Sigma(r)$ are given in Gorski (1988). From a computational point of view we find more convenient to adopt the equivalent formulae given by Peebles (1993), [his Eqs. (21.72) and (21.74)]:

$$\Pi(r) = \frac{2}{3}\left(\frac{H_0 \Omega_0^{0.6}}{\sqrt{2}}\right)^2 \left[\frac{J_5(r)}{r^3} + K_2(r)\right] \qquad (8)$$

$$\Sigma(r) = \left(\frac{H_0 \Omega_0^{0.6}}{\sqrt{2}}\right)^2 \left[\frac{J_3(r)}{r} - \frac{J_5(r)}{3r^3} + \frac{2}{3}K_2(r)\right] \qquad (9)$$

with $J_n(r)$ and $K_n(r)$ given by:

$$J_n(r) = \int_0^r dr \xi(r) r^{n-1}, \quad K_n(r) = \int_r^\infty dr \xi(r) r^{n-1}$$

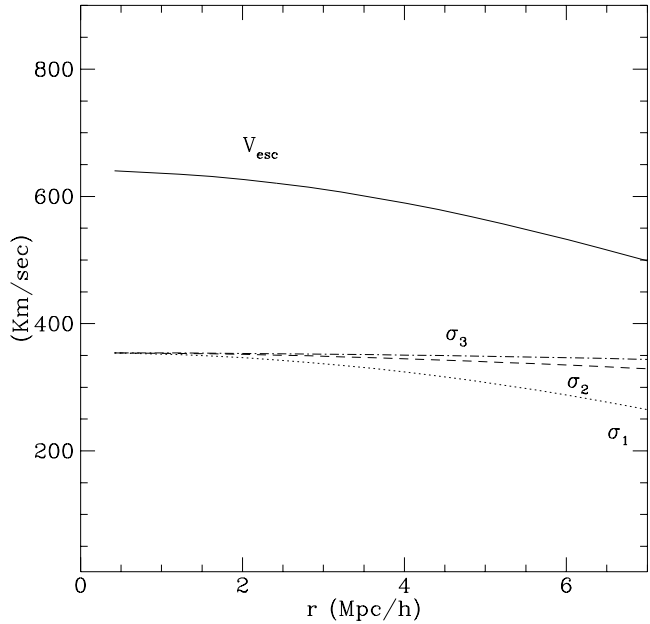

**Fig. 1.** Eigenvalues of the velocity correlation tensor for a standard CDM model with $b_8 = 1.1$, a filtering radius $R_f = 5$ Mpc and $\nu = 1.2$.

From BBKS [1986, see their eq. (A3)] we have: $\langle \nu x \rangle = \gamma$. The other averages appearing in the various terms of the matrix $M_{ij}$ are given by:

$$\langle v_i \nu \rangle = -\frac{H_0 \Omega_0^{0.6}}{r^3 \sigma_0} r_i \int_0^r dr' r'^2 \xi(r'), \qquad (10)$$

$$\langle v_i \eta_j \rangle = -\frac{}{r^3} r_i r_j \left\{ r\xi(r) - \int_0^r dr'\xi(r') \right\}, \qquad (11)$$

$$\langle v_i x \rangle = \frac{H_0 \Omega_0^{0.6}}{r \sigma_2} r_i \frac{d\xi}{dr}. \qquad (12)$$

(see the Appendix A for more details on the derivation of these quantities).

From the distribution function given in Eq. 5 we can obtain the average velocity distribution around peaks of arbitrary height by imposing the condition $\boldsymbol{\eta} = 0$ and integrating over all values $x \geq 0$. Eventually we obtain:

$$f_{\rm pk}(\boldsymbol{v} \mid \nu) = \frac{(\alpha/4\pi^2)}{(\prod_{i=1}^3 \lambda_i)^{1/2}} \exp\left(-\sum_{i=1}^3 \frac{\{v_i - \nu\langle v_i\nu\rangle\}^2}{\lambda_i}\right) \qquad (13)$$

where we have defined:

$$\alpha = \frac{\sqrt{\pi}}{2} erfc\left[\frac{\nu\langle \nu x\rangle}{\sqrt{\lambda_4}}\right]$$

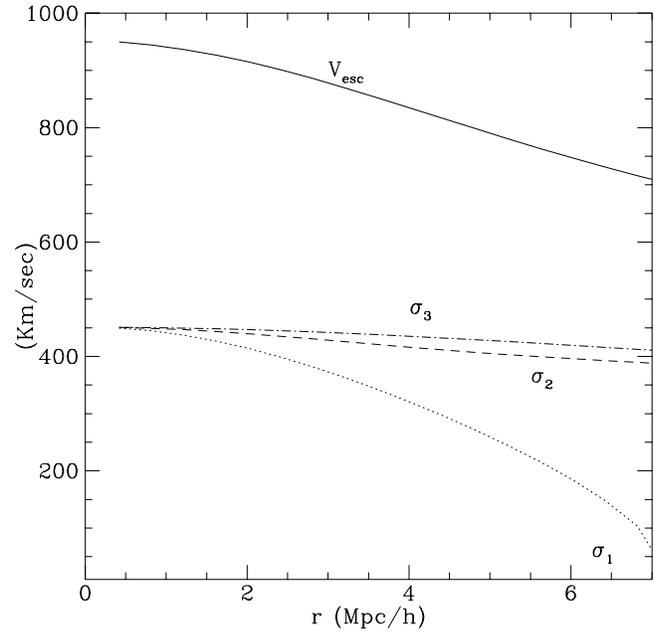

**Fig. 2.** Same as Fig. 1 for $\nu = 2$ and for $R_f = ...$ such that the total mass at a given distance of 10Mpc is $M \approx 2.725 \times 10^{15} M_\odot$.

From Eq. (13) the physical meaning of the eigenvalues $\lambda_i, i = 1,...3$ is clear: they are the components of the peculiar velocity dispersions along the parallel ($\lambda_1$) and perpendicular ($\lambda_{2,3}$) w.r.t. the radial direction of a density peak. In Figs. 1 and 2 we plot the profiles of these quantities for two configurations having approximately the same total mass ($M \approx 2.725 \times 10^{15} h M_\odot$) at a reference distance of $10 h^{-1}$Mpc (comoving units). This choice of mass and total cluster masses as derived from gravitational lensing observations in nearby clusters (Kneib et al. 1995; Waxman & Miralda-Escudé 1995). For example, in Abell 1689 (Tyson & Fischer 1995) a mass M ≈ $2.7 \times 10^{15} h^{-1} M_\odot$ is found within $R_A = 1.5 h^{-1}$ Mpc. The collapse factor of a density shell reaching an overdensity $\bar\delta_f$ is given by $x_c = [1 + \bar\delta_f]^{1/3}$ (Ryden & Gunn 1987), and for $\bar\delta_f = 300$ we get $x_c \approx 6.6$. Thus, the final radius of a shell at $10 h^{-1}$ Mpc is approximately equal to one Abell radius, for values of the central overdensity typical of rich Abell clusters. High $\nu$, i.e. more centrally condensed density peaks have a larger spatial gradient of the velocity dispersion, as one can intuitively expect. It is however interesting to note the considerable *quantitative* difference among the central values of the velocity dispersions: about 100 Km/sec. This effect clearly suggests a broad variance in the magnitude of the expected bulk flows.

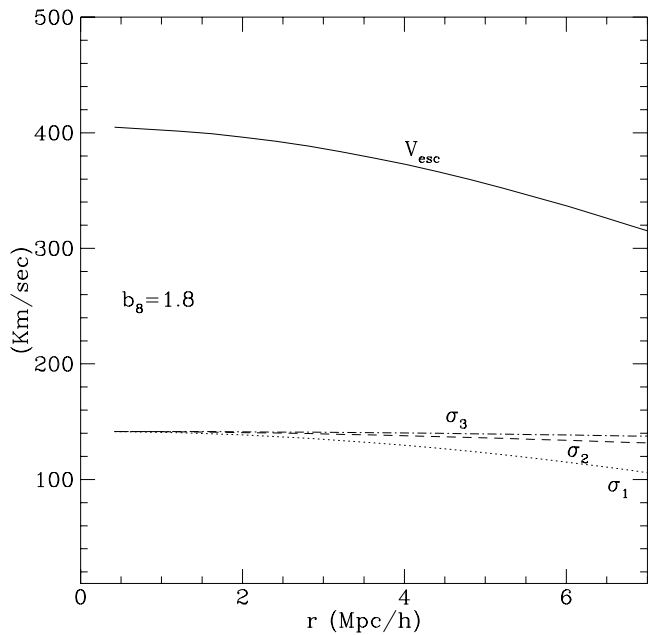

**Fig. 3.** Same as Figure 1 for $b_8 = 1.8$.

*2.2. Dispersion profiles*

The three components of the velocity dispersion attain nearly the same value in the central regions of the peaks, within the filtering radius (in Fig.1 $R_f = 5\ h^{-1}$ Mpc while in Fig.2 $R_f \simeq 2\ h^{-1}$ Mpc). The velocity dispersion $\lambda_1$ in the radial direction decreases faster with the distance $r$ from the center of the peak than the tangential velocity dispersions $\lambda_2$ and $\lambda_3$. In fact, at larger radii $r$, the radial component is sensibly lower than the tangential com-of the clusters are predominantly non-radial. For objects with the same mass, the onset of non-radial motions happens at shorter radii when the filtering radius is smaller and consequently the central height of the peak is higher (see Fig.2).

In the very central regions of the peaks, $r \lesssim 0.5 R_f$, there is no net difference among the three components of the velocity dispersions and they are approximately constant meaning that this region is dominated by an isotropic velocity distribution.

The region contained within the filtering radius does not show a net predominance of radial motions. As our considerations refer to protostructures still in the early phases of their non-linear evolution, this region can be probably considered as the one that will collapse in the final non-linear configuration through an infall process.

Regions of the protostructures at $r \gtrsim R_f$ contain mainly non-radial motions. In these outer regions the fate of the infalling material could be strongly dependent on the amount of tangential velocity dispersion relative to the radial, inward acceleration. This can be also seen by looking at the equation of motion of a spherically spherically symmetric mass distribution with density $n(r)$:

$$\frac{\partial}{\partial t} n(r) \langle v_r \rangle = -n(r) \frac{\partial \langle v_r \rangle}{\partial t} - \frac{\partial}{\partial r} n(r) \langle v_r^2 \rangle +$$

$$n(r) \frac{\langle v_t^2 \rangle - 2 \langle v_r^2 \rangle}{r} \qquad (14)$$

From the structure of the right hand side of equation (11), it is easy to see that high tangential velocity dispersions, $\langle v_t^2 \rangle \gtrsim 2 \langle v_r^2 \rangle$ may alter the infall patterns.

Some recent numerical simulations (Evrard & Crone 1992), show that in the infall regime the velocity is primarily radial. These results were obtained using numerical codes with limited dynamical range; in fact, we believe that high resolution simulations are needed to determine the detailed structure of the velocity field in clusters of galaxies. But based on our results we are lead to conclude that in the early stages of the cluster evolution the velocity distribution is not prevalently radial, even tough the strong non-linear evolution tends to modify the velocity distribution of the infalling material. However, as clusters of galaxies – being the largest and youngest gravitationally bound systems in the universe – show a broad variance of dynamical states, the debate on the structure of their velocity distribution is still open.

Finally, note that from Figs. 1-2 typical values for the velocity dispersions at $1\ h^{-1}$ Mpc are in the range $350 - 450$ Km/sec, i.e. well within the $1\sigma$ values found by Marzke et al. (1995): $540 \pm 180$ km/sec. These quantities are not easily comparable to the pairwyse velocity correlations measured by the latter authors and by others (e.g. Davis & Peebles 1983), because pairs of galaxies can be found within region of different average overdensity. Even admitting that the galaxy pairs selected in a survey sample fairly the (average) peculiar velocity field of the regions to which they belong, one should expect anyhow a rather large *intrinsic* uncertainty at any scale because of the *intrinsic* variance of the peculiar dispersion profiles. We will deal with these problems in a future paper (Antonuccio-Delogu et al. 1995).

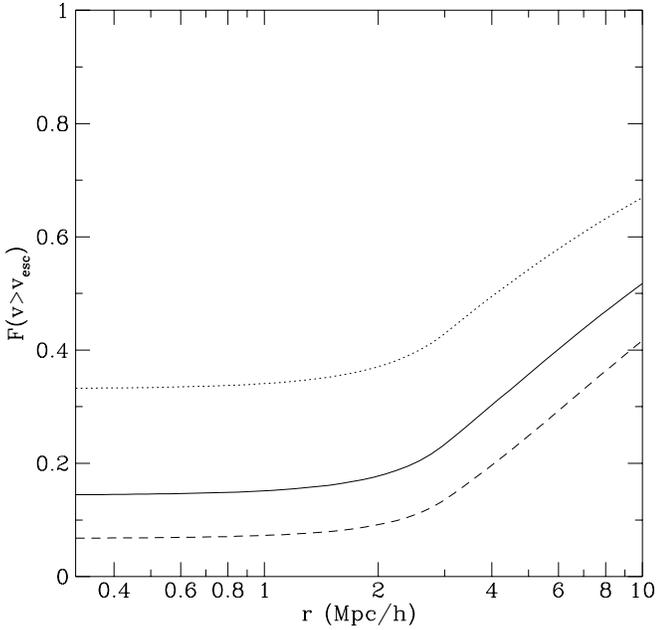

**Fig. 4.** Spatial variation of $F(|v| \geq v_{vesc})$ in the case of a top-hat perturbation of radius $R = 1.5h^{-1}$ Mpc. The three curves refer to values of the velocity dispersion, $\sigma_v = 800$ km/s (dashed curve), $\sigma_v = 1000$ km/s (continuous curve) and $\sigma_v = 1500$ km/s (dotted curve). Here $h = 0.5$.

## 3. Fraction of gravitationally unbound matter

The previous considerations allow us to evaluate the fraction of matter that is gravitationally bound (in a statistical sense) to the protostructure. A simple criterion to define a gravitationally bound, infalling shell is to compare the velocity dispersion at distance $r$ with the escape velocity $v_{\rm esc}(r) = (2 |\Phi(r)|)^{1/2}$ at the same distance from the center of the protostructure.

A non zero fraction of unbound matter could be expected even in the case of a uniform system with a Maxwellian velocity distribution, $f(v) \propto exp(-v^2/2\sigma_v^2)$, characterized by a uniform velocity dispersion $\sigma_v^2$. In fact, in the case

Mpc (the Abell radius), the fraction

$$F(v > v_{\rm esc}) = \frac{\int_{|v| \geq v_{\rm esc}} d^3\boldsymbol{v} f(\boldsymbol{v})}{\int d^3\boldsymbol{v} f(\boldsymbol{v})} \tag{15}$$

takes values $\approx 0.08 \div 0.33$ for $\sigma_v = 800 \div 1500$ km/s (see Fig. 4). This fraction rises up to values $0.4 \div 0.7$ at distances $r \sim 5h^{-1}$ Mpc for the same values of the velocity dispersion.

For the case of a density peak, the fraction of *unbound* matter reads:

$$F(v > v_{esc}) = \frac{\int_{|v| \geq v_{\rm esc}} d^3\boldsymbol{v} f_{\rm pk}(\boldsymbol{v}, \nu)}{\int d^3\boldsymbol{v} f_{\rm pk}(\boldsymbol{v}, \nu)} \tag{16}$$

We compute this quantity for a spherically symmetric density peak in a standard CDM model. We have been able to reduce the integral in the numerator of Eq. 16 to a one dimensional integral which we evaluate numerically: the interested reader will find the details of this calculation in Appendix B.

In Fig.5 we show the result in the case of peaks with different heights subtending the same mass as given in the preceding section ($M_{10h^{-1}Mpc} = 2.725 \times 10^{15} M_\odot$).

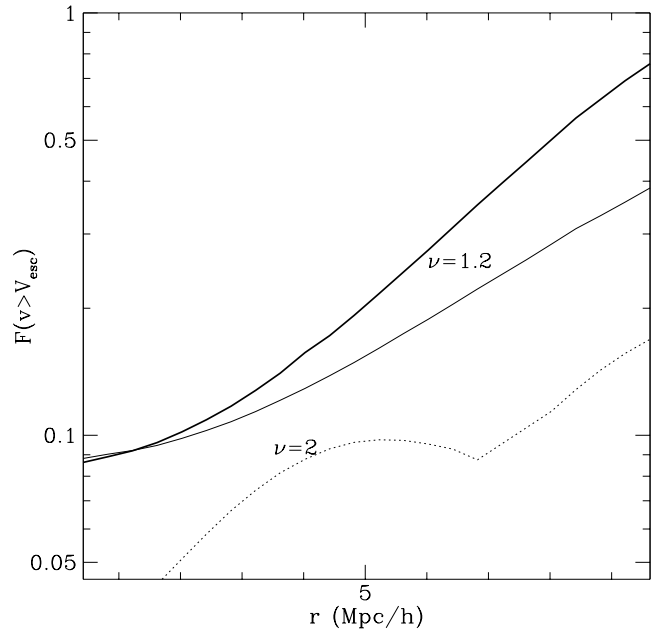

**Fig. 5.** Spatial variation of $F(|v| \geq v_{vesc})$. The thick solid curve is for $b_8 = 1.6$, the other two for $b_8 = 1.1$. The total mass for each curve is the same as in Figs. 1-3.

For a relatively low peak $\nu = 1.2$ (which is quite frequent at cluster scales) we obtain that the fraction of unbound clumps within $R_f = 5h^{-1}$ Mpc is $F(v > v_{esc}) \approx$ at $\sim 2R_f = 10h^{-1}$ Mpc. This fraction lowers for higher $\nu$ peaks, as the potential of these protostructure is deeper and hence the value of $v_{esc}$ is systematically higher. We found, in fact that, for a peak with $\nu = 2$ and $R_f = 2h^{-1}$ Mpc the fraction $F(v > v_{esc}) \approx 0.04 \div 0.07$ within $R_f$, and rises up to values $\gtrsim 0.1$ at $2R_f$.

## 4. Conclusions

In this paper we have been mainly concerned with the structure of the velocity field that is found around the local maxima of the density fluctuation field at early times. We have derived analytically the velocity distribution function around a density peak starting from the approach of BBKS, originally used to derive the density profile of the peak in a Gaussian random field.

From the analysis of the three components of the velocity correlation tensor, we found that peaks show the presence of a nearly isotropic region in the very central regions (at scales $r \approx 0.5 R_f$). Non-radial motions, instead, become dominant in the outer regions of the protostructures at $r \gtrsim R_f$.

As the region of the protostructures that develops infall is found probably at $R \lesssim R_f$, we are lead to the conclusion that in the early stages of the protostructure evolution there is not a coherent infall of matter during the collapse of the structure. This situation might involve a modification of the collapse dynamics of such structures, at least in their early phases.

In the standard CDM scenario the flatness of the power spectrum at high wavenumbers produces an high fraction of clumps on small scales populating the external regions of the clusters. According to the previous results, these clumps can have high velocities (momenta), even larger than the average escape velocity from the structure. As the escape velocity is a measure of the cluster gravitational potential, the frequency of clumps that can be statistically of high momenta (those we call "unbound") is larger in the outer regions of the protostructure. In fact, we evaluated that the fraction of unbound clumps within the filtering radius of the clusters is not negligible: we found values $F(v > v_{esc}) \sim 0.05 \div 0.4$ for quite common peaks, with central heights in the range $\nu \sim 1.2 \div 2$.

These findings indicate that there exist a large fraction of small mass clumps with statistically high momenta in the peripherical regions, $r \gtrsim R_f$, of the protoclusters. During the collapse of the cluster, and also during the subsequent secondary infall, these high velocity clumps can exchange momentum with the collapsing or infalling clumps and generate a drag force (produced by dynamical friction) onto the collapsing material that acts to delay their collapse time scales (see Antonuccio-Delogu & Colafrancesco 1994, Colafrancesco et al. 1995). The set of high momentum clumps provides a field of stochastic force generators that intervenes in modifying the equation of a delay of the collapse time particularly affecting the external shells of the clusters. This reflects eventually in a collapse delay of the overall structure. Such considerations support and strengthen the conclusions of our previous work (Colafrancesco et al. 1995) concerning the dynamical effect of small-scale substructure on the collapse of clusters of galaxies.

*Acknowledgements.* The investigation of V.A.-D. was partly supported by the European Community-Human Capital and Mobility network program *ULySSES*. V.A.-D. thanks B.J.T. Jones for the very helpful discussion and the support for his visit at TAC.

## A. Appendix: Velocity field correlations

The ensemble-averaged vector products appearing in Eqs. (6) are linear combinations of the following averages:

$$\langle x^2 \rangle, \langle x\nu \rangle, \langle \nu^2 \rangle, \langle \boldsymbol{v}\nu \rangle, \langle \boldsymbol{v}\boldsymbol{\eta} \rangle, \langle \boldsymbol{\eta}\boldsymbol{\eta} \rangle \tag{A1}$$

From BBKS, eq. (A3) we have that:

$$\langle x^2 \rangle = 1, \langle x\nu \rangle = \gamma, \langle \nu^2 \rangle = 1 \tag{A2}$$

where: $\gamma \equiv \sigma_1^2/\sigma_2\sigma_0$ and $\sigma_i$ is the $i$-th momentum of the variance of the fluctuation field:

$$\sigma_i^2 \equiv \frac{1}{2\pi^2} \int P(k) k^{2(i+1)}$$

Here $P(k)$ is the *filtered* power spectrum.

Remembering the definition: $\nu \equiv \delta(0)/\sigma_0$ and the definition of correlation function:

$$\xi(r) \equiv \langle \delta(r) \delta(0) \rangle \tag{A3}$$

and adopting Eq. 1 for the peculiar velocity in the linear regime, we immediately find:

$$\langle \boldsymbol{v}\nu \rangle = -\frac{H_0 \Omega_0^{0.6}}{r^3} \cdot \frac{\boldsymbol{r}}{\sigma_0} \int_0^r dr' r'^2 r'^2 \xi(r') \tag{A4}$$

Let us now evaluate the quantity $\langle \boldsymbol{v}\boldsymbol{\eta} \rangle$. The components of this tensor are given by:

$$\langle v_i \eta_j \rangle = -\frac{H_0 \Omega_0^{0.6}}{r^3} r_i \int_0^r dr' r'^2 \langle \bar{\delta}(r') \eta_j \rangle,$$

where $\eta_j$ are the components of the gradient of the density field evaluated at the center of the peak density profile:

$$\eta_j = \left.\frac{d\bar{\delta}(r)}{dr_j}\right|_{r=0} \tag{A5}$$

Performing an integration by parts and recalling the definition of correlation function (eq. A3) we obtain:

$$\begin{aligned}\langle v_i \eta_j \rangle &= -\frac{H_0 \Omega_0^{0.6}}{r^3} r_i \int_0^r dr' r'^2 \langle \bar{\delta}(r') \left.\frac{\bar{\delta}(r)}{dr_j}\right|_{r=0} \rangle \\ &= -\frac{H_0 \Omega_0^{0.6}}{r^3} r_i \int_0^r dr' r'^2 \frac{r_j}{r'} \frac{d\xi(r')}{dr'} = \\ &= -\frac{H_0 \Omega_0^{0.6}}{r^3} r_i r_j \left\{ r\xi(r) - \int_0^r dr' \xi(r') \right\}\end{aligned} \tag{A6}$$

Our purpose is to evaluate the integral in the numerator of Eq. 16 for the probability distribution given in Eq. 13. We make a change of coordinates in velocity space: $(v_1, v_2, v_3) \to (v_r, v_T, \psi)$ where we have defined: $v_1 =: v_r$, $v_2 =: v_T \cos(\psi)$, $v_3 =: v_T \sin(\psi)$. It is easy to check that the Jacobian is given by:

$$\left| \frac{\partial (v_1, v_2, v_3)}{\partial (v_1, v_T, \psi)} \right| = |v_T| \tag{B1}$$

The probability distribution $f_{\rm pk}(\boldsymbol{v}, \nu)$ is normalized, so the numerator in Eq. 16 can be expressed as:

$$\int_{|v| \geq v_{\rm esc}} d^3 \boldsymbol{v} f_{\rm pk}(\boldsymbol{v} \mid \nu) = 1 - \int_{|v| < v_{\rm esc}} d^3 \boldsymbol{v} f_{\rm pk}(\boldsymbol{v} \mid \nu)$$

and we will now give a formula for the integral in the right-hand side of this equation.

With the coordinate transformation given above we find, after some simple algebra:

$$\int_{|v| < v_{\rm esc}} d^3 \boldsymbol{v} f_{\rm pk}(\boldsymbol{v} \mid \nu) =$$

$$= \frac{(\alpha/4\pi^2)}{\left(\prod_{i=1}^3 \lambda_i\right)^{1/2}} \int_{-v_{\rm esc}(r)}^{v_{\rm esc}(r)} dv_r \exp\left[-\frac{(v_r - \nu \langle v\nu\rangle r)^2}{\lambda_1}\right] \cdot$$

$$\cdot \int_0^{\sqrt{v_{\rm esc}^2(r) - v_r^2}} dv_T \, v_T \exp\left[-\frac{v_T^2}{2}\left(\frac{1}{\lambda_2} + \frac{1}{\lambda_3}\right)\right] \cdot$$

$$\cdot \int_0^{2\pi} d\psi \exp\left[-\frac{v_T^2}{2}\left(\frac{1}{\lambda_2} - \frac{1}{\lambda_3}\right) \cos(2\psi)\right] \tag{B2}$$

With the help of Eq. (9.6.16) from Abramowitz & Stegun (1972) the integral over the angular direction becomes:

$$\int_0^{2\pi} d\psi \exp\left[-\frac{v_T^2}{2}\left(\frac{1}{\lambda_2} - \frac{1}{\lambda_3}\right) \cos(2\psi)\right] =$$

$$= 2\pi I_0 \left( \frac{v_T^2}{2} \left[ \frac{1}{\lambda_2} - \frac{1}{\lambda_3} \right] \right) \tag{B3}$$

where $I_0$ denotes the modified Bessel function of zero order. We now define the function:

$$F\left(\eta^2, \gamma, \beta\right) = \int_0^{\eta^2} dz \, e^{-\beta z} I_0(\gamma z) \tag{B4}$$

so that Eq. B2 becomes:

$$\int_{|v| < v_{\rm esc}} d^3 \boldsymbol{v} f_{\rm pk}(\boldsymbol{v}, \nu) = \frac{\alpha}{4\pi \prod_{i=1}^3 \lambda_i} \cdot$$

$$\int_{-v_{\rm esc}(r)}^{v_{\rm esc}(r)} \exp\left[-\frac{(v_r - \nu \langle v\nu\rangle r)^2}{\lambda_1}\right] F\left(\eta^2(r), \gamma, \beta\right) \tag{B5}$$

$$\eta^2(r) = v_{\rm esc}^2(r) - v_T^2, \tag{B6}$$

$$\gamma = -\frac{1}{2}\left(\frac{1}{\lambda_2} - \frac{1}{\lambda_3}\right), \quad \beta = \frac{1}{2}\left(\frac{1}{\lambda_2} + \frac{1}{\lambda_3}\right) \tag{B7}$$

After having evaluated numerically the function $F(\eta_2, \gamma, \beta)$ we can perform numerically the integral (Eq. B5). This allows us to calculate the fraction of unbound particles.